\documentclass[%
reprint,
superscriptaddress,
amsmath,
amssymb,
aps,
prb,
]{revtex4-2}
\usepackage{graphicx}
\usepackage{dcolumn}
\usepackage{bm}
\usepackage[hidelinks]{hyperref} 

\newcommand{\iu}{\ensuremath{\mathrm{i}}}

\newcommand{\fig}[1]{{Fig.~{\ref{#1}}}}

\usepackage{mathtools}
\DeclarePairedDelimiter\bra{\langle}{\rvert}
\DeclarePairedDelimiter\ket{\lvert}{\rangle}
\DeclarePairedDelimiterX\braket[2]{\langle}{\rangle}{#1\,\delimsize\vert\,\mathopen{}#2}

\newcommand{\varket}[1]{\lvert {#1} \rangle}
\newcommand{\varev}[1]{\langle {#1} \rangle} 
\newcommand{\varip}[2]{\langle {#1} | {#2} \rangle}

\newcommand{\ev}[1]{\left\langle {#1} \right\rangle} 
\newcommand{\ip}[2]{\left\langle {#1} \middle| {#2} \right\rangle}
\newcommand{\op}[2]{{\ket{#1}\bra{#2}}}
\newcommand{\Tr}{\mathrm{Tr}}
\newcommand{\dv}[1]{\frac{\mathrm{d}}{\mathrm{d}{#1}}}
\newcommand{\pdv}[1]{\frac{\partial}{\partial{#1}}}
 
\newcommand{\mel}[3]{\bra{#1}{#2}\ket{#3}}
\newcommand{\dd}[1]{\mathrm{d}{#1}}
\newcommand{\abs}[1]{\left\lvert {#1} \right\rvert}
\newcommand{\qand}{\quad\mathrm{and}\quad}
\newcommand{\comm}[2]{\left[#1,#2\right]}
\newcommand{\order}[1]{\mathcal{O}(#1)}
\newcommand{\qtya}[1]{{\left( {#1} \right)}}
\newcommand{\qtyb}[1]{{\left[ {#1} \right]}}

\newcommand{\be}{\begin{equation}}
\newcommand{\ee}{\end{equation}}

\newcommand{\mc}[1]{\mathcal{#1}}

\newcommand{\Fig}[1]{Figure~\ref{#1}}

\newcommand{\eq}[1]{Eq.~\eqref{#1}}
\newcommand{\eqs}[2]{Eqs.~\eqref{#1} and \eqref{#2}}
\newcommand{\Eq}[1]{Equation~\eqref{#1}}

\newcommand{\stn}[1]{Sec.~\ref{#1}}

\renewcommand{\Re}{\mathrm{Re}}

\newcommand{\bea}{\begin{eqnarray}}
\newcommand{\eea}{\end{eqnarray}}
 
\newcommand{\ba}{\begin{array}}
\newcommand{\ea}{\end{array}}

\newcommand{\bl}{\begin{flalign}}
\newcommand{\enl}{\end{flalign}}

\begin{document}

\title{Bexcitonics: quasiparticle approach to open quantum dynamics}%

\author{Xinxian Chen}
\email{xchen106@ur.rochester.edu}
\affiliation{Department of Chemistry, University of Rochester, Rochester, New York 14627, United States}

\author{Ignacio Franco}
\email{ignacio.franco@rochester.edu}
\affiliation{Department of Chemistry, University of Rochester, Rochester, New York 14627, United States}
\affiliation{Department of Physics, University of Rochester, Rochester, New York 14627, United States}

\date{\today}

\begin{abstract}

{We develop a quasiparticle approach to capture the dynamics of open quantum systems coupled to bosonic thermal baths of arbitrary complexity based on the Hierarchical Equations of Motion (HEOM).
This is done by generalizing the HEOM dynamics and mapping it into that of the system in interaction with a few bosonic fictitious quasiparticles  that we call bexcitons.
Bexcitons arise from a decomposition of the bath correlation function into discrete features. Specifically, bexciton creation and annihilation couple the auxiliary density matrices in the HEOM.  
The approach provides a systematic strategy to construct exact quantum master equations that include the system-bath coupling to all orders even for non-Markovian environments. Specifically, by introducing different metrics and representations for the bexcitons it is possible to straightforwardly generate different variants of the HEOM, demonstrating that all these variants share a common underlying quasiparticle picture.
Bexcitonic properties, while unphysical, offer a coarse-grained view of the correlated system-bath dynamics and its numerical convergence. For instance, we use it to analyze the instability of the HEOM when the bath is composed of underdamped oscillators and show that it leads to the creation of highly excited bexcitons. 
The bexcitonic picture can also be used to develop more efficient approaches to propagate the HEOM. As an example, we use the particle-like nature of the bexcitons to introduce  mode-combination of bexcitons {in both number and coordinate representation} that uses the multi-configuration time-dependent Hartree to efficiently propagate the HEOM dynamics.}

\end{abstract}

\maketitle
\section{Introduction}

A central challenge in the physical sciences is to accurately capture the quantum dynamics of atoms, molecules and other quantum systems when they interact with quantum thermal environments~\cite{Breuer2002,Schlosshauer2007,Nielsen2011,May2011,Tanimura2020,Cygorek2022,Landi2022}.
This is needed, for example, to develop better organic solar cells~\cite{Popp2019}, understand vital processes such as photosynthesis~\cite{Cao2020}, and to advance quantum technologies for computing, sensing and communication~\cite{Koch2016,Koch2022}. Several strategies have been developed to follow this open quantum system dynamics through quantum master equations (QMEs) that implicitly capture the influence of the quantum bath on the system.  In particular, important progress has been made for bosonic environments~\cite{Tanimura1990, Makri1995, Makri1995a, Vega2015, Strathearn2018, Lambert2019, Tamascelli2019, Tanimura2020, Kim2021}. These environments are ubiquitous because any quantum environment can be mapped into a collection of bosons provided the system-bath interaction can be captured to second order in perturbation theory~\cite{Feynman1963, Caldeira1983, Caldeira1993}, and this situation is common in the condensed phase~\cite{Suarez1991, Makri1995} where system-bath interactions are diluted over a macroscopic number of degrees of freedom.

\begin{figure}[tb]
    \centering
    \includegraphics[width=\linewidth]{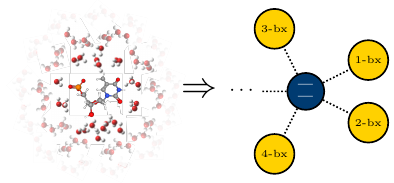}
    \caption{The open quantum dynamics can be exactly mapped to that of the quantum system interacting with a few bexcitons, fictitious quasiparticles arising from distinct features of the bath correlation function.}
    \label{fig:diagram}
\end{figure}

{One of the most powerful numerical methods to simulate the open quantum dynamics is the numerically exact Hierarchical Equations of Motion (HEOM)~\cite{Tanimura1989,Tanimura1990, Shi2009,Ishizaki2009,Ikeda2020,Tanimura2020}.
The original HEOM was developed as a simulation method for open quantum system coupled with a Drude--Lorentz bath~\cite{Tanimura1990} which yields a bath correlation function that decays exponentially in time. Since then it has been extended to many other types of baths, including Brownian and discrete vibrational baths~\cite{Liu2014} which yield bath correlation functions that are oscillatory. 
The HEOM is analogous in form to the stochastic Liouville equation but with additional dissipation terms that are needed to correctly capture the energy flow between system and environment.~\cite{Tanimura2006} The HEOM exactly captures the dynamics of systems interacting with thermal harmonic environments, admits arbitrary time dependent in the system, and can be extended to fermionic environments~\cite{Schinabeck2016}.  Since the approach avoids invoking perturbation theory in the system-bath interaction, it goes beyond all methods based on perturbative expansions such as the Lindblad and Redfield equations~\cite{Lindblad1976,Redfield1965,Lidar2019}.}

{Several variants of the HEOM have been proposed such as the extended HEOM~\cite{Tang2015, Xu2022, Xu2023}, generalized HEOM~\cite{Ikeda2020, Ikeda2022} and the hierarchical Schr{\"{o}}dinger equations of motion~\cite{Nakamura2018, Tokieda2020}. 
These HEOM variants result from adopting different decompositions of the bath correlation function that lead to distinct, but closely related, quantum master equations of varying computational complexity. }

{In this paper, we develop a quasiparticle approach to capture the dynamics of open quantum systems based on the HEOM. This is done by generalizing the HEOM dynamics and mapping it into that of the system in interaction with a few {collective bath excitations} or bexcitons (see \fig{fig:diagram}). The bexcitons are fictitious quasiparticles that arise from a decomposition of the bath correlation function into distinct features, and are created and destroyed as the system decoheres. Specifically, bexciton creation and annihilation couple the auxiliary density matrices in the HEOM.  While in quantum master equations the environment's dynamics is not followed explicitly, in this bexcitonic picture it is captured in a coarse-grained way offering additional tools to understand the system--bath entanglement and numerical convergence of the method.}

{The bexcitons admit representation in an arbitrary basis and offer flexibility in their metric, thus the bexcitonic equations represent a whole class of HEOM-like equations of motion. In fact, using this approach, we show how to straightforwardly recover different variants of the HEOM by choosing different metrics and basis representations for the bexcitons, and develop new ones.  Because all these HEOM variants are seen to be specific realizations of the same bexcitonic equations of motion, when converged they yield the same dynamics. However, the convergence property and numerical stability may vary because of the different errors introduced in the truncation of bexcitonic space in different realizations.}


{The bexcitonic approach offers advantages both in terms of interpretation of the open quantum dynamics and in the development of efficient numerical propagation schemes. As an example of the former,   we discuss the instability of the HEOM by illustrating the bexcitonic properties when the bath is composed of underdamped oscillators and show that it leads to the creation of highly excited bexcitons. As an example of the latter, we show how the bexcitonic picture can be used to develop more efficient approaches to propagate the HEOM. Specifically, we exploit the particle-like nature of the bexcitons to introduce  mode-combination of bexcitons that uses the multi-configuration time-dependent Hartree to efficiently propagate the HEOM dynamics.}

{The structure of this paper is as follows. In \stn{stn:theory}, we derive the bexcitonic quantum master equation, discuss the emerging bexcitonic picture and its relation with existing methods. In \stn{stn:numerics}, we exemplify the bexcitonic dynamics in position and number representation, and discuss the influence of the metric on numerical convergence.  In \stn{stn:applications}, we use the bexcitonic picture to investigate the numerical instability of the HEOM and to develop more efficient propagation schemes.  We summarize our main findings in  \stn{stn:conclusions}.}

\section{Theory}
\label{stn:theory}

As customary, we decompose the Hamiltonian of a quantum system in interaction with an environment $H = H_{\text{S}}(t) + H_{\text{SB}} + H_{\text{B}}$,
into the system  $H_{\text{S}}(t)$,  the bath $H_{\text{B}}$, and  their interaction  $H_{\text{SB}}$. The system can be subject to arbitrary time-dependence, such as that introduced by light-matter interactions. 
For bosonic baths, $H_{\text{B}} = \sum_{j} \omega_j  a_{j}^\dagger a_{j}$
where $\omega_j$ is the frequency of the $j$-th  mode and $a_j^\dagger$ its raising and $a_j$ its lowering operator ($[a_j, a_i^\dagger] = \delta_{ij}$). 
For simplicity, we focus on $H_{\text{SB}} = Q_{\text{S}} \otimes X_{\text{B}}$ with only one coupling term but the final results are readily extendable to many coupling terms.
Here $Q_{\text{S}}$ is a system's operator and $X_{\text{B}} =\sum_j \qtya{ g_{j}a_{j}^\dagger + g_{j}^{\star}a_j }$ a collective bath coordinate with $g_j$ coupling strength to the $j$-th mode. Throughout we use atomic units where $\hbar=1$.

We begin from the exact dynamical map $\mathcal{V}(t)$ of the system's reduced density matrix $\rho_{\text{S}}(t) = \mathcal{V}(t) \rho_{\text{S}}(0)$  at time $t$ from initial state $\rho_{\text{S}}(0)$~\cite{Ishizaki2009}. For convenience we use the notation $A^>B = AB$ and $A^<B = BA^\dagger$ for the ordering of matrix multiplications, and $A^{\times} = A^> - A^<$  and $A^{\circ} =  A^> + A^<$ for the symmetric and anti-symmetric super-operator generated from $A$.  Dynamical maps require the overall density matrix $\rho(t)$ to be initially in a separable state 
$\rho(0) = \rho_{\text{S}}(0) \otimes \rho_\text{B}^{\text{eq}}$. We take $\rho_\text{B}^{\text{eq}}= e^{-\beta H_{\text{B}}} / {\mc{Z}}$  to be the thermal density matrix of the bath, where $\beta = 1/k_\text{B}T$,  $T$ the temperature, ${\mc{Z}}= \Tr_{\text{B}}e^{-\beta H_{\text{B}}}$ the bath partition function, and $\Tr_\text{B}$ denotes a trace over bath degrees of freedom. While the dynamics of $\rho(t)$ is unitary, the dynamics of $\rho_{\text{S}}(t) = \Tr_{\text{B}}\rho(t)$ is non-unitary and satisfies~\cite{Ishizaki2009}
\begin{equation}\label{eq:prop}
    {\tilde{\rho}_{\text{S}}(t)}= \mathcal{T} \tilde{\mathcal{F}}(t,0) \rho_{\text{S}} (0),
\end{equation}
where $\mathcal{T}$ is the time-ordering operator,
\begin{equation}\label{eq:if}
    \tilde{\mathcal{F}}(t,0) = e^{- \int_0^t\dd{s} \tilde{Q}^{\times}_{\text{S}}(s) \int_0^s \dd{u} \qtya{C(s-u) \tilde{Q}_{\text{S}}(u)}^{\times}},
\end{equation}
and $C(t) = \Tr{\tilde{X}_\text{B}(t) \tilde{X}_\text{B}(0) \rho_\text{B}^{\text{eq}}}$ is the bath correlation function (BCF). In writing \eq{eq:prop} we have adopted the interaction picture of $H_0(t) = H_{\text{S}}(t) + H_{\text{B}}$,  where $\tilde{O}(t) = \mathcal{T} e^{ \iu\int_0^t H_0(t') \dd{t'}} O(t) \mathcal{T} e^{-\iu\int_0^t H_0(t') \dd{t'}}$. 
\Eq{eq:if} provides a formal solution to the open quantum dynamics at all temperatures and to all orders in the system-bath interaction. As seen, $C(t)$ contains \emph{all} the information needed to capture the influence of the bath on $\rho_{\text{S}}(t)$. 

\subsection{Identifying dynamical features of the bath correlation function}

To make this formal solution computationally tractable, 
we decompose $C(t)$ and its conjugate $C^{\star}(t)$ as 
\begin{equation}\label{eq:bcf}
    C(t) = \sum_{k=1}^K c_k \psi_k(t), \qand C^{\star}(t) = \sum_{k=1}^K \bar{c}_k \psi_k(t),
\end{equation}
where $\{\psi_k(t)\}$ is a complex \emph{basis} and $c_k$, $\bar{c}_k$ are time-independent complex expansion coefficients. 
Each component of the basis defines a \emph{feature} of the bath and is required to satisfy
\begin{equation}\label{eq:bcf-basis}
\begin{aligned}
    \dv{t} \psi_k(t) =  \sum_{j=1}^K \gamma_{kj}\psi_j(t), &\qand \psi_k(0) = 1.
\end{aligned}
\end{equation}
The first condition guarantees that $\{\psi_k(t)\}$ spans a function space that contains both  $C(t)$ and its time-derivative, as needed for dynamics. The second one reflects that physical systems have non-zero quantum fluctuations.
The dimension $K$ of this basis defines the number of bath features. 
As discussed below, this decomposition is general but not unique.

{The decomposition of $C(t)$ into features is a necessary step in all variants of HEOM as it is needed to make the open quantum dynamics practical. Where strategies differ is in the specifics of the decomposition, as that leads to different master equations with different computational complexities and numerical properties. ~\cite{Tanimura1989, Ishizaki2009, Tang2015, Ikeda2020, Xu2022}}

Any basis satisfying \eq{eq:bcf-basis} can be used to decompose the BCF into features. 
We now show a systematic, albeit not unique, way to do this that demonstrate \eq{eq:bcf} is general, and that yields features satisfying $\gamma_{kk'}= 0$ for $k\ne k'$. The structure of the bath is captured by its spectral density $J(\omega) =  \sum_j \abs{g_j}^2 \delta(\omega - \omega_j)$ ($\omega > 0$), a quantity that summarizes the frequencies of the environment and its interaction strength to the system. The BCF is related to $J(\omega)$ through~\cite{Callen1951,May2011}
\begin{equation}\label{eq:fdt}
    C(t) = \int_{-\infty}^{+\infty}\mathcal{J}(\omega) f_\text{BE}(\beta\omega) e^{-\iu\omega t}\dd{\omega},
\end{equation}
where $\mathcal{J}(\omega) = \sum_j \abs{g_j}^2 \qtyb{\delta(\omega - \omega_j) - \delta(\omega + \omega_j)}$ is an odd extension of $J(\omega)$ and 
$f_{\text{BE}}(\beta\omega) = (1 - e^{-\beta\omega})^{-1}$ is the Bose-Einstein distribution.
We evaluate \eq{eq:fdt} using the residue theorem through analytical continuation and expanding $f_\text{BE}(\beta\omega)$ through a Pad{\'{e}}~\cite{Hu2010} or Matsubara~\cite{Zheng2009} schemes (see also Refs.~\cite{Cui2019,Zhang2020,Xu2022}). 
In both cases, 
\begin{equation}\label{eq:cexpansion}
    \begin{aligned}
    C(t) &= - 2 \pi \iu \sum_{i} {\mathop {\operatorname {Res}}_{z=\zeta_i}}[\mathcal{J}(z)] f_{\text{BE}}(\beta \zeta_i) e^{-\iu \zeta_i t}\\
    &\quad- 2 \pi \iu \sum_{j} {\mathop {\operatorname {Res}}_{z=\xi_j}}[f_{\text{BE}}(z)] \mathcal{J}({\xi_j}/{\beta}) e^{-\iu (\xi_j/\beta)t},
    \end{aligned}
\end{equation}
where $\{\zeta_i\}$ are the first order poles of $\mc
{J}(\omega)$ and  $\{\xi_j\}$ those of  $f_{\text{BE}}(\beta \omega)$ (in the lower-half complex plane). 
These expansions satisfy \eq{eq:bcf} with each term defining a feature.  
The expansion of $f_{\text{BE}}(\beta \omega)$ leads to exponentially decaying $\psi_k(t)$ (as its Pad{\'{e}} and Matsubara expansions have purely imaginary poles). 
By contrast, the poles of the spectral density can lead to other types of bath correlations.

As two important cases, we now isolate this dynamics for the Drude--Lorentz (DL) and Brownian environments which are the basic models for condensed phase environments~\cite{Mukamel1995, Gustin2023} though \eq{eq:bcf} and can be used for other types of physical spectral densities~\cite{Kim2022}.
For simplicity in presentation, we focus on the high temperature limit case where only the poles from $J(\omega)$ are considered.
However, the approach is general and the computations presented do not make this simplification.

The DL spectral density~\cite{Caldeira1981, Grabert1988, Tanimura1990}
\begin{equation}
    J(\omega) = \frac{2\lambda}{\pi} \frac{\omega_{\text{c}}\omega}{\omega^2 + \omega_{\text{c}}^2}
\end{equation}
models Ohmic environments with cutoff frequency $\omega_{\text{c}}$ and reorganization energy $\lambda$. 
In this case,  $C(t) = c_1 e^{- \omega_{\text{c}} t}$ decays exponentially on a time scale $\omega_{\text{c}}^{-1}$, and $c_1 = \lambda \omega_{\text{c}} \qtya{\cot(\beta\omega_{\text{c}}/2)-\iu}$ describes the coupling strength between the system and the bath.
Contrasting with \eq{eq:bcf} we see that there is only one feature needed to describe this dynamics as $\gamma_1= -\omega_{\text{c}}$ and it is inherently dissipative. Features that arise from low-temperature corrections to $f_\text{BE}(\beta \omega)$ are also of this kind.

The Brownian spectral density
\begin{equation}
    J(\omega) = \frac{4\lambda}{\pi} \frac{\eta \omega_0^2 \omega}{(\omega^2 - \omega_0^2)^2 + 4\eta^2\omega^2}
\end{equation}
describes a discrete harmonic oscillator of natural frequency $\omega_0$ damped at a rate $\eta$~\cite{Garg1985,Liu2014}. 
In this case, the BCF exhibits oscillations of frequency $\omega_1 = \sqrt{\omega_0^2 - \eta^2} > 0$ that decay at a rate $\eta$ as 
$C(t) = c_1 e^{\gamma_{1} t}  + c_2 e^{\gamma_{2} t}$, where $\gamma_{1} = - \eta + \iu\omega_1$, $\gamma_{2} = - \eta - \iu\omega_1$
and $c_1 = c_+$, $c_2 = c_-$ with the system--bath coupling strength determined by $c_{\pm} = \lambda\omega_1 \qtya{1 + \eta^2/\omega_1^2} \qtya{\coth({\beta\qtya{\omega_1 \pm \iu\eta}}/{2}) \mp 1}/2$
. 
Thus, at least two features are needed to capture this system-bath dynamics.   

\subsection{An exact quantum master equation for open quantum dynamics}

Using the decomposition of the BCF \eq{eq:bcf}, the propagator in \eq{eq:if} can be separated into contributions by different bath features as
\begin{equation}\label{eq:propagator}
    \tilde{\mathcal{F}}(t,0) =\prod_{k=1}^{K} 
    e^{- \int_0^t\dd{s} \tilde{Q}^{\times}_{\text{S}}(s) \tilde{f}_k(s, 0)},
\end{equation}
where
$\tilde{f}_k(s,0) =  c_k\tilde{\theta}_k^>(s,0) - \bar{c}_k \tilde{\theta}_k^<(s,0)$ and
$\tilde{\theta}_k(s, 0) =  \int_0^{s} \tilde{Q}_{\text{S}}(u)\psi_k(s-u) \dd{u}$. 
To exactly capture the open quantum dynamics, we need to take into account how each bath feature influences the system's dynamics through $\tilde{f}_k$. 
For this, we define a hierarchy of auxiliary density matrices as
\begin{align}\label{eq:adm-def} 
    \tilde{\varrho}_{\vec{n}}(t) \equiv
        {\mathcal{T}  
        \qtya{\prod_{k=1}^{K}
        \frac{\tilde{f}^{n_k}_k (t, 0)}{Z_{k}(n_k) \sqrt{n_k!}} }
        \tilde{\mathcal{F}}(t,0)}
        \rho_{\text{S}}(0).
\end{align}
Here, $Z_{k}(n_k) =  \prod_{m = 1}^{n_k} {z_{k, m}} $ for $n_k > 0$ and $Z_{k}(n_k) = 1$ for $n_k = 0$, and ${z_{k, m}}$ are non-zero $c$-numbers which we refer as the metric of feature $k$.
The index $\vec{n}$ indicates a multi-dimensional index $\vec{n} = (n_1, \cdots, n_k, \cdots,  n_K)$ with $n_k=0,\ 1,\ 2,\ \ldots$ and the series runs \emph{ad infinitum}. 
The physical system's density matrix ${\rho}_{\text{S}}(t) = {\varrho}_{\vec{0}}(t)$ is located at $\vec{n} = \vec{0} \equiv (0, \cdots, 0)$ 

We define an extended density operator (EDO) as a collection of these auxiliary density matrices. 
We arrange these matrices as a vector of matrices 
$\ket{\tilde{\varrho}(t)} =\sum_{\vec{n}} \tilde{\varrho}_{\vec{n}}(t) \ket{\vec{n}} $ in a basis $\{\ket{\vec{n}}\equiv \ket{n_1}\otimes\cdots\otimes \ket{n_k}\otimes\cdots\otimes\ket{n_K} \}$
such that $\tilde{\varrho}_{\vec{n}}(t) = \ip{\vec{n}}{\tilde{\varrho}(t)}$. 
We define the creation $\alpha^\dagger_k$ and annihilation $\alpha_k$ operators associated to the $k$-th bath feature such that
\begin{equation}
    \hat{\alpha}^\dagger_k \ket{{n_k}} = \sqrt{n_k + 1} \ket{{n_k+1}},\    \hat{\alpha}_k \ket{{n_k}} = \sqrt{n_k} \ket{{n_k-1}}, 
\end{equation}
with $[\hat\alpha_k, \hat\alpha_{k'}^\dagger] = \delta_{k,k'}$, $\hat{n}_k=\hat\alpha_{k}^\dagger\hat\alpha_{k}$ and $\hat{n}_k\ket{n_k} = n_k\ket{n_k}$.
We also define the metric operator for the $k$-feature as $\hat{z}_k \ket{n_k} =  z_{k,n_k} \ket{n_k}$,
which implies that $\comm{\hat{z}_k}{\hat{n}_k} = 0$ as they admit a common eigenbasis. 

Next, we determine the equation of motion of the EDO, $\ket{\tilde{\varrho}(t)}$, by direct differentiation of \eq{eq:adm-def}. 
The system-bath interaction couples the different auxiliary density matrices as the dynamics of $\tilde{\varrho}_{\vec{n}}(t)$ is coupled to 
$\tilde{\varrho}_{\vec{n}+\vec{1}_k}(t) = \tfrac{1}{\sqrt{n_k+1}}\bra{\vec{n}}\hat\alpha_k \ket{\tilde{\varrho}(t)} $,  $\tilde{\varrho}_{\vec{n}-\vec{1}_k}(t) = \tfrac{1}{\sqrt{n_k}} \bra{\vec{n}}\hat\alpha_k^\dagger \ket{\tilde{\varrho}(t)} $ and $\tilde{\varrho}_{\vec{n}-\vec{1}_{k}+\vec{1}_{k'}}(t) = \tfrac{1}{\sqrt{n_k(n_{k'}+1)}}\bra{\vec{n}} \hat\alpha_{k}^\dagger \hat\alpha_{k'} \ket{\tilde{\varrho}(t)}$ for $k \neq k'$, where $\vec{n} \pm \vec{1}_k$ increases/decreases $n_k$ by one, leaving all other indexes in $\vec{n}$ intact. 
In the Schr\"odinger picture, 
\begin{equation}\label{eq:eom}
     \pdv{t} \ket{{\varrho}(t)} =\qtya{ - {\iu} {H}^{\times}_{\text{S}}(t) + \sum_{k=1}^{K}\mathcal{D}_k} \ket{{\varrho}(t)},
\end{equation}
where 
\begin{equation}
\begin{aligned}\label{eq:dissipator}
           \mathcal{D}_k 
       &= \gamma_{kk} {\hat{\alpha}}^\dagger_k {\hat{\alpha}}_{k} +
       \sum_{k'\neq k} \gamma_{k k'} \hat{z}_{k}^{-1} {\hat{\alpha}}^\dagger_k {\hat{\alpha}}_{k'} \hat{z}_{k'} 
       \\        &\quad
       +  \qtya{
        c_k {Q}_\text{S}^> - \bar{c}_k {Q}_\text{S}^< }
        \hat{z}_{k}^{-1} \hat{\alpha}^\dagger_k 
       - {Q}^{\times}_\text{S}
        \hat{\alpha}_k\hat{z}_{k}
\end{aligned}
\end{equation}  
are the dissipators $\mathcal{D}_k$ associated with the $k$-th bath feature.
\Eq{eq:eom} together with the initial condition $\ket{\varrho(0)} = \rho_{\text{S}}(0)\varket{\vec{0}}$ exactly specifies the open quantum dynamics.  
A detailed derivation is included in the Appendix~\ref{sec:derivation}.

\Eq{eq:eom} leads to a bexcitonic picture of the open quantum dynamics and can be used to construct practical {HEOM-like} exact quantum master equations (EQME).  It shows that each basis $\{\psi_k(t)\}$ used to capture the BCF, metric $\{\hat{z}_k\}$ and representation of $\hat{\alpha}_k$,  leads to a distinct EQME. These equations can appear to be very different but originate from the same \eq{eq:eom}.  \Eq{eq:eom} defines a class of EQMEs.

\subsection{Bexcitonic picture}
We now discuss how a bexcitonic picture emerges from \eq{eq:eom}. 
We associate $\ket{\vec{n}}$ with the creation of \emph{bexcitons}, with respect to vacuum $\varket{\vec{0}}$.
Specifically, we associate a bexciton of label $k$, a $k$-bexciton, for each feature of the bath $k$. The state $\ket{\vec{n}}$ corresponds to a situation in which $n_k$ $k$-bexcitons have been created for each $k$. In this picture, $\hat{\alpha}_k^\dagger$ creates and  $\hat{\alpha}_k$ destroys a $k$-bexciton. 
The commutation relation between $\hat{\alpha}_k$ and $\hat{\alpha}_k^\dagger$ dictates that the algebra for the bexcitons are bosonic.  
While the bath can be macroscopic, only $K$ effective bexcitons are needed to capture the relevant component that influences the system. 
Thus, the bexcitons offer a coarse-grained, but still exact, view of the correlated non-Markovian system-bath dynamics to all orders in $H_{\text{SB}}$.

The dissipators $\{\mc{D}_k\}$ in \eq{eq:dissipator} describe the bexcitonic dynamics and their interaction with the system. 
At initial time, the system-bath density matrix is separable and there are no bexcitons. As the composite system evolves toward a stationary state, bexcitons are created and destroyed. The first term in \eq{eq:dissipator}  describes the decay (for $\gamma_{kk}<0$), oscillations (for purely imaginary $\gamma_{kk}$),  or both, of the bexcitons. The second, describes possible bexciton-bexciton interactions.  The third, corresponds to the creation of bexcitons due to system-bath interaction while the last term leads to bexciton annihilation. The bexcitons do not keep track of the orders in a perturbative expansion in $H_{\text{SB}}$ as many orders can contribute to a given bexcitonic population.  

The number of bexcitons $K$ needed to accurately describe the dynamics increases as the complexity of the spectral density grows (which requires more $\{\zeta_i\}$) and with decreasing temperature (which requires more $\{\xi_j\}$) as showed in the decomposition of BCF in \eq{eq:cexpansion}. 
Equation~\eqref{eq:cexpansion} also shows that, bexciton--bexciton interactions are zero since the time dependence is in the exponentials leading to $\gamma_{kk'} = \gamma_{kk}\delta_{kk'}$ in \eq{eq:bcf-basis}. 
Thus, from this point on, without loss of generality we take $\gamma_{kk'} = \delta_{kk'}\gamma_{k}$.

\Eq{eq:eom} exactly maps the open quantum dynamics to the system-bexciton dynamics.   While the system's dynamics is common to all maps, the bexcitonic one is not. For this reason, the bexcitons are unphysical quasiparticles and bexcitonic properties should only be seen as a way to monitor the dynamics and numerical convergence of a particular EQMEs in the class.

To quantify bexcitonic properties it is necessary to specify an inner product for the EDOs. Given two EDOs, $\varket{\varrho^{(1)}}$ and $\varket{\varrho^{(2)}}$, we define their inner product as
$\varip{\varrho^{(1)}}{\varrho^{(2)}} \equiv \sum_{\vec{m},\vec{n}} \varev{\varrho^{(1)}_{\vec{m}}, \varrho^{(2)}_{\vec{n}}} \ip{\vec{m}}{\vec{n}}$,
where $\ev{A, B}\equiv \Tr(A^\dagger B) = \sum_{ij}A_{ij}^{\star}B_{ij}$ is the Hilbert--Schmidt inner product between matrices $A$ and $B$.
In this way, the expectation value of a Hermitian bexcitonic operator, $\bra{\varrho(t)} \hat{O} \ket{\varrho(t)} = \sum_{\vec{m},\vec{n}} \varev{\varrho_{\vec{m}}, \varrho_{\vec{n}}} \bra{\vec{m}}\hat{O}\ket{\vec{n}},$ is scalar and real. For instance, the population of the $k$-bexciton 
$\ev{n_k} = {\bra{\varrho} \hat{n}_k \ket{\varrho}} = \sum_{\vec{n}} n_k \ev{\varrho_{\vec{n}},\varrho_{\vec{n}} }$, and the system purity $\mathcal{P} \equiv \Tr_{\text{S}} \rho^2_{\text{S}} = \ev{\varrho_{\vec{0}}(t),\varrho_{\vec{0}}(t)} = \bra*{\varrho} P_0 \varket{\varrho}$ with $P_0=\op{\vec{0}}{\vec{0}}$.

\subsection{Bexcitonic representations}
We now develop two useful general forms for the EQMEs. For this, we need to specify the representation of the bexcitonic operators $\{\hat{\alpha}_k, \hat{\alpha}^\dagger_k\}$.
The most immediate way to represent the bexcitons is in their occupation number representation $\{  \ket{\vec{n}}\}$:
\begin{widetext}
 \begin{equation}\label{eq:heom}
    \partial_{t} \varrho_{\vec{n}}(t) 
    = -{\iu}\comm{{H}_\mathrm{S}(t)}{\varrho_{\vec{n}}(t)} 
    + \sum_{k=1}^K\qtya{n_k \gamma_{k} {\varrho}_{\vec{n}}(t) 
    -  z_{k,n_k+1}\sqrt{n_k + 1} \comm{{Q}_\mathrm{S}}{{\varrho}_{\vec{n} + \vec{1}_k}(t)}  
    + \frac{\sqrt{n_k}}{z_{k,n_k}} 
    \qtya{c_k {Q}_\mathrm{S} {\varrho}_{\vec{n} - \vec{1}_k}(t) - \bar{c}_k {\varrho}_{\vec{n} - \vec{1}_k}(t) {Q}_\mathrm{S} }
    }
    .
\end{equation} 
This equation recovers the HEOM, see Sec.~\ref{sec:w-heom}.
Thus, \eq{eq:eom} can be seen as a generalization of the HEOM strategy. 

The bexcitons can also be represented in position, $\vec{x}= (x_1, \cdots, x_K)$, where $x_k$ is the position of the $k$-bexciton, by letting $\hat{\alpha}^\dagger_k \to  \qtya{x_k - \partial_{x_k}}/\sqrt{2}$ and $\hat{\alpha}_k \to \qtya{x_k + \partial_{x_k}}/\sqrt{2}$ such that $[\hat\alpha_k, \hat\alpha_{k'}^\dagger] = \delta_{k,k'}$.
In this case,
\begin{equation}\label{eq:pde-BME}
    \partial_{t} \varrho(\vec{x}, t) = 
    - {\iu} \comm{{H}_{\text{S}}(t)}{{\varrho}(\vec{x}, t)} 
    + \sum_{k=1}^K \qtya{\frac{\gamma_{kk}}{2}  \qtya{ x_k^2 - \partial^2_{x_k} - 1}{\varrho}(\vec{x}, t)
    - \iu {Q}_{\text{S}} \qtya{ g_k^- x_k - g_k^+ \partial_{x_k}}{\varrho}(\vec{x}, t)
    + \iu \qtya{ \bar{g}_k^- x_k - \bar{g}_k^+\partial_{x_k}}
    {\varrho}(\vec{x}, t){Q}_{\text{S}}
    },
\end{equation}
\end{widetext}
where $g^\pm_k = \iu(c_k z_{k}^{-1} \pm z_{k}) / \sqrt{2}$,
$\bar{g}^\pm_k = \iu(\bar{c}_k z_{k}^{-1} \pm z_{k}) / \sqrt{2}$.
For simplicity, in writing \eq{eq:pde-BME} we have taken $\hat{z}_k =z_k \hat{1}$. In this form, the initial condition is 
$\varrho(\vec{x}, 0) = \rho_{\text{S}}(0) G(\vec{x})$ and the system's density matrix $\rho_{\text{S}}(t) = \int \varrho(\vec{x}, t)G(\vec{x}) \dd[K]{\vec{x}}$, where $G(\vec{x}) = \langle\vec{x}|\vec{0}\rangle ={\pi}^{-K/4} \prod_{k=1}^{K} {e^{-x_k^2/2}}$. 
{\Eq{eq:pde-BME} is} closely related to the collective bath coordinate method~\cite{Ikeda2022}{ by adopting different metric}.

While \eqs{eq:heom}{eq:pde-BME} have vastly different forms they are seen to be specific representations of \eq{eq:eom}. The approach opens the way to systematically develop different representations for the EQMEs including number, position and momentum $\vec{p}$ (that can be obtained from \eq{eq:pde-BME} by letting $x_k\to p_k$, and $g_k^{\pm} \to -\iu g_k^{\mp}$) and even mixed representations where different representations are used for each bexciton.

{\subsection{Recovering standard HEOM equations}}\label{sec:w-heom}

We now show that \eqs{eq:eom}{eq:dissipator} generalize the HEOM in the sense that the standard HEOM equations are seen to emerge as specific cases.
If we take the specific metric operator $\hat{z}_k = \iu \hat{n}_k^{-1/2}$, and use the number representation, we obtain
\begin{widetext}
\begin{equation}  
    \pdv{t} \varrho_{\vec{n}}(t) 
    = -\iu \comm{{H}_\mathrm{S}}{\varrho_{\vec{n}}(t)} + \sum_k n_k \gamma_{k} {\varrho}_{\vec{n}}(t)
    - \iu  \comm{{Q}_\mathrm{S}}{{\varrho}_{\vec{n} + \vec{1}_k}(t)} 
    - \iu\sum_k n_k \qtya{
            c_k {Q}_\mathrm{S}{\varrho}_{\vec{n} - \vec{1}_k}(t) - \bar{c}_k {\varrho}_{\vec{n} - \vec{1}_k}(t){Q}_\mathrm{S} }
    , 
\end{equation}
which is exactly the standard HEOM~\cite{Shi2009, Liu2014, Tanimura2020} for both Drude--Lorentz and Brownian environments. In turn, if we let $\hat{z}_{k} = \iu\sqrt{\abs{c_k}}$, we get 
\begin{equation}\label{eq:si-heom-og}  
        \pdv{t} \varrho_{\vec{n}}(t) 
    = -\iu  \comm{{H}_\mathrm{S}}{\varrho_{\vec{n}}(t)} 
    + \sum_k n_k \gamma_{k} {\varrho}_{\vec{n}}(t) 
    - \iu\sum_k {\sqrt{({n_k + 1}){\abs{c_k}}}} \comm{{Q}_\mathrm{S}}{{\varrho}_{\vec{n} + \vec{1}_k}(t)} 
    - \iu\sum_k\sqrt{\frac{n_k}{\abs{c_k}}} \qtya{
            c_k {Q}_\mathrm{S}{\varrho}_{\vec{n} - \vec{1}_k}(t) - \bar{c}_k {\varrho}_{\vec{n} - \vec{1}_k}(t){Q}_\mathrm{S}}
    .
\end{equation}
\end{widetext}
This equation coincides with the main result of HEOM with scaling in Ref.~\onlinecite{Shi2009} [Eq.~(6)] if we further restrict the bath to the Drude--Lorentz case where $c^{\star}_k = \bar{c}_k$, and the correlation function is fitted to a series of decaying exponentials. 

These examples show that many variants of the HEOM can be developed starting from \eqs{eq:eom}{eq:dissipator} simply by changing the metric operator $\hat{z}_k$ and the representation. All these variants arise from a common bexcitonic picture.

\subsection{Relation to existing methods in literature}

Equations \eqref{eq:eom} and \eqref{eq:dissipator} define a class of exact quantum master equation when the decomposition of the BCF into features is exact.
In this section, we contrast the choices that define the method with other strategies in the literature.

{\subsubsection{Bath correlation function decomposition}}

{We first note that a wide range of variants of the HEOM use a decomposition of the BCF that can be cast in the form in \eqs{eq:bcf}{eq:bcf-basis}. For example, the form of the BCF decomposition that includes both exponential and oscillatory terms was  suggested in Ref.~\onlinecite{Liu2014} and used to exemplify the dynamics generated by Drude-Lorentz and Brownian oscillator environments. \Eq{eq:fdt} shows how the exponential and oscillatory terms in such a decomposition can be developed systematically through a Pad\'e or Matsubara scheme that supposes that there are only first-order poles in the spectral density $\mathcal{J}(\omega)$ and the thermal distribution function $f_{\text{BE}}(\beta\omega)$. Recently, Xu \emph{et al.}~\cite{Xu2022} demonstrated that the first-order pole decomposition is general using a rational function approach, emphasizing the general applicability of \eqs{eq:bcf}{eq:bcf-basis}. Even more general decompositions to evaluate \eq{eq:fdt} that include higher-order poles, such as those developed by Ikeda and Scholes~\cite{Ikeda2020}, can also be employed to develop a bexcitonic picture.  However, in light of Ref.~\onlinecite{Xu2022} they are not formally necessary. However, they may lead to more efficient versions of the HEOM.}

{In decomposing the BCF into features it is also not necessary to use the residue theorem to evaluate \eq{eq:fdt}. For example, the extended HEOM~\cite{Tang2015} uses a decomposition of the BCF using harmonic oscillator wavefunctions $\psi_k = H_k(\alpha t) \exp\qtyb{-\alpha^2 t^2/2}$ as a basis, where $H_k$ are Hermite polynomials. This basis satisfies the dynamics in \eq{eq:bcf-basis} but not the initial conditions as $H_k(0) =0$ for odd $k$. This basis leads to interacting bexcitons and, as it decays in time with a Gaussian envelope, it is not efficient to capture weakly damped vibrations as that requires many terms in the decomposition. However, this is an example of a possible basis that can be made to satisfy \eq{eq:bcf-basis} even when it does not arise from the residue theorem.}

{Overall, the fact that the decomposition of the BCF in \eqs{eq:bcf}{eq:bcf-basis} is widely used in different HEOM variants, indicates that the bexcitonic quasiparticle  structure is common to all of them.}

{\subsubsection{Quasiparticle views of open quantum dynamics}}

{Another quasiparticle view of the open quantum dynamics that yields the standard HEOM for bosonic environments is the Dissipaton Equations of Motion~\cite{Yan2014,Li2023}. In this view, the dissipatons are physical collective bath degrees of freedom. Each dissipaton yields a specific term in the decomposition of the bath correlation function into features. By contrast, the bexcitons are fictitious quasiparticles that account for the algebraic structure of the quantum master equation, and that enable unifying different variants of the HEOM into a single framework.  These two quasiparticles definitions  are distinct and compatible as they enter in different steps of the derivation of HEOM-style quantum master equation.
The dissipatons have been extended to fermionic environments and non-linear system-bath couplings~\cite{Xu2018}, suggesting that it is also possible to do a similar analysis for the bexcitons. The advantage of the bexcitons is that it enables one to straightforwardly develop variants of the HEOM by adopting different metrics and representations for the creation and annihilation operators, as needed for specific applications.}

{For instance, when the position basis is adopted for the bexcitons, it yields equations that are closely related to the recently proposed collective bath coordinate method~\cite{Ikeda2022} but with a tunable metric that provides additional flexibility. In Ref.~\onlinecite{Ikeda2022}, in the light of \eqs{eq:eom}{eq:dissipator}, the authors effectively do a transformation from the number basis to position representation. However, the underlying mathematical structure of such a transformation had not been sufficiently clarified as needed for systematic progress. Through \eqs{eq:eom}{eq:dissipator}, it becomes straightforward to systematically change representation and metric to develop EQMEs.}

{In turn, Ref.~\onlinecite{Xu2023} also introduces a raising and lowering operators to capture the bath dynamics from the free-pole-HEOM~\cite{Xu2022}. However, in Ref.~\onlinecite{Xu2023} these operators act in Liouville space instead of Hilbert space. This choice results in twice the number of indexes for the auxiliary density matrices for a given BCF decomposition, which squares the computational complexity. Further, thus far, the computational efficiency and stability of the method have not been demonstrated. }


{A different strategy to develop a quasiparticle decomposition of the open quantum dynamics is the pseudo-mode method~\cite{Lambert2019,Landi2022,Anto2023}. This method introduces a small number of ``unphysical'' harmonic modes with dissipative terms, and enlarge the system of interest to include these pseudo-modes, such that the ``unphysical'' model has equivalent BCF compared with the actual physical model. As such, it is best suited for underdamped Brownian environments and requires further techniques for Ohmic environments~\cite{Somoza2019}. By contrast, the bexcitons can exactly treat both overdamped and underdamped environments, Markovian and non-Markovian dynamics, weak and strong system-bath correlations, in a unified form by propagating the dynamics in bexcitonic space.}  From the geometric point of view, the pseudo-mode method and the bexcitonics are defined in very different mathematical spaces.
The pseudo-mode method  constructs an enlarged dissipative system as $\mathsf{S} \otimes {\mathsf{P}}$, where $\mathsf{S}$ denotes the Liouville space of the system and $\mathsf{P}$ is the {Liouville space} of the pseudo-modes, and the density matrix of the system of interest is calculated by {tracing out} the degrees of freedom of the pseudo-modes $\mathsf{P}$.
By contrast, in the bexcitionics dynamics the EDO is in space $\mathsf{S} \otimes \mathsf{Q}$ where $\mathsf{Q}$ is the {Hilbert space} of bexcitons (as opposed to a Liouville space).
The reduced density matrix of the system is recovered by {projection} (as opposed to tracing out) onto the vacuum state $\varket{\vec{0}}$ of $\mathsf{Q}$. 
These differences make the bexcitonics very different from the pseudo-mode approach and provide a distinct point of view to understand the open quantum system dynamics.

\section{Numerical implementation}
\label{stn:numerics}

To demonstrate the utility of \eq{eq:eom} in simulating open quantum dynamics, we computationally implemented it in both number and position representation. As in HEOM, the ladder of states for each $n_k$ needs to be truncated at a given $N_k$ that defines the \emph{depth} of the $k$-bexciton in number representation, a quantity that needs to be increased until convergence. 
In position representation, we employed two forms (Sinc-DVR and Sine-DVR) of the discrete variable representation (DVR)~\cite{Colbert1992, Harris1965} which provide an efficient grid representation~\cite{Littlejohn2002}.
In this case, the depth $N_k$ is determined by the number of grid points in the allowed range $(-\tfrac{L_k}{2}, \tfrac{L_k}{2})$ for $x_k$. 
The overall space complexity of \eq{eq:eom} for a $M$-state system and $K$ bath features is $\order{M^2N^K}$. 
\Eq{eq:eom} was implemented using the PyTorch package~\cite{Paszke2019} for efficient CPU and GPU computation and is available on GitHub~\footnote{X.~Chen. BEX: A general python package to simulate open quantum systems. Available at \texttt{https://github.com/vINyLogY/bex}}.


As a specific model, consider a qubit with $H_{\text{S}} = ({\Delta}/2) \sigma_z + V \sigma_x$, where $\sigma_z = \op{\text{e}}{\text{e}} - \op{\text{g}}{\text{g}}$ 
and $\sigma_x = \op{\text{e}}{\text{g}} + \op{\text{g}}{\text{e}}$ are Pauli operators and $\ket{\text{g}}$ and $\ket{\text{e}}$ the qubit levels.
The dynamics starts from a pure state $\rho_{\text{S}}(0) = \op{\psi}{\psi}$ describing a superposition of qubit levels $\ket{\psi} = \qtya{\ket{\text{g}} + \ket{\text{e}}}/ \sqrt{2}$.
Suppose the characteristic energy of the system is $E \neq 0$.
At $t=0$ the system is coupled through $Q_{\text{S}} = \sigma_z$ to a  bath at temperature  $k_\mathrm{B}T = 0.209 E$ described by a DL (with $\lambda = 0.2 E$ and $\omega_{\text{c}} = 0.1 E$) or Brownian spectral density (with $\lambda = 0.2 E$, $\omega_1 = E$, and $\eta = 0.05 E$).  

As a metric, we use $\hat{z}_k = z_k \hat{1}$ where $z_k = \iu\sqrt{\Re\, c_k}$ for the DL case and $z_k = \iu\sqrt{\Re (c_k + \bar{c}_k)/2}$ for Brownian. {In this section}, we find that $K=3$ and a depth of $N_k = 10$ for the number-basis provide converged dynamics. The grid basis requires a larger $N_k = 40$ (for $L_k=40$).

\subsection{Pure-dephasing dynamics}

\begin{figure}[tb]
    \centering
    \includegraphics[width=\linewidth]{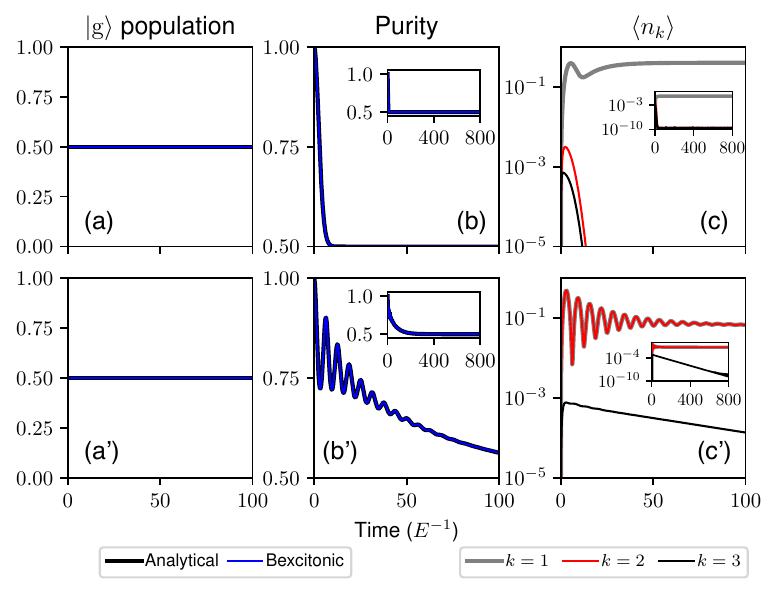}
    \caption{Pure-dephasing ($V = 0$) dynamics of a qubit initially in a superposition state $\ket{\psi} = \qtya{\ket{\text{g}} + \ket{\text{e}}}/ \sqrt{2}$ interacting with a thermal bath as described by \eq{eq:eom} using the number representation. The plots show (a-a') population of $\ket{\text{g}}$; (b-b') state purity; (c-c') bexcitonic population for $k=1,\ 2,\ 3$. The upper panels (a-c) correspond to a DL bath, while (a'-c') are those generated by a Brownian bath. The insets show the asymptotic dynamics. The black  line in (a-b, a'-b') corresponds to the analytical solution.}
    \label{fig:dynamics-pd}
\end{figure} 

Consider first the case in which $V=0$ and $\Delta = E$.
In this case the dynamics is pure-dephasing as $\comm{H_\text{S}}{H_{\text{SB}}}=0$, and independent of $\Delta$ as there is no relaxation.
This limit of the dynamics admits an analytical solution~\cite{Gustin2023, Schlosshauer2007} that we now use to test the bexcitonic formalism in number representation [\eq{eq:heom}].
\Fig{fig:dynamics-pd} shows the dynamics of the (a-a') population of $\ket{\text{g}}$, (b-b') the qubit purity $P(t)$ and (c-c') the $k$-bexciton population. 
The top panels (a-c) are for the DL bath while the bottom panels (a'-c') for the Brownian bath. 
As shown, the bexcitonics exactly reproduces the analytical results.
In the DL case, the purity decays monotonically as expected for a system interacting with a macroscopic environment and settles at $\mc{P} = 1/2$ which corresponds to the maximally mixed state.
By contrast for the Brownian environment the purity exhibits an oscillatory dynamics before decay to $1/2$.
These oscillations are due to changes in system--bath entanglement as the Brownian environment oscillates. The purity asymptotically stays at $\mc{P} = 1/2$ as there are no relaxation process at play.

With respect to the bexcitons, initially, the $k$-bexciton population is $\ev{n_k}=0$. Upon time evolution the population of all three considered bexcitons initially increases as system-bath interact.
For DL, the $k=1$ bexciton is the high-temperature term in \eq{eq:cexpansion} and $k=2,\ 3$ are the low temperature corrections.
For Brownian $k=1$ and $2$ are the high-temperature terms and $k=3$ is a low temperature correction.
For Brownian, there is a population degeneracy of the ($k=1,\ 2$) bexcitons needed to describe $J(\omega)$. 

In this pure-dephasing case, the high-temperature bexcitons reach steady state with non-zero $\ev{n_k}$ at the long-time limit.
By contrast, the low temperature correction terms $\ev{n_k}$ go to zero after the initial excitation.
The time required for the bexciton population to reach the steady state coincides with the time needed for the system to reach the maximally mixed state with $\mathcal{P} = 1/2$.
Therefore, the bexciton population reflects the entanglement between the system and the bath.
The non-zero bexciton population at the final steady state reflects the non-separable system-bath state at equilibrium.

\subsection{Relaxation dynamics of biased ($\Delta \neq 0$) qubit}
\begin{figure}[tb]
    \centering
    \includegraphics[width=\linewidth]{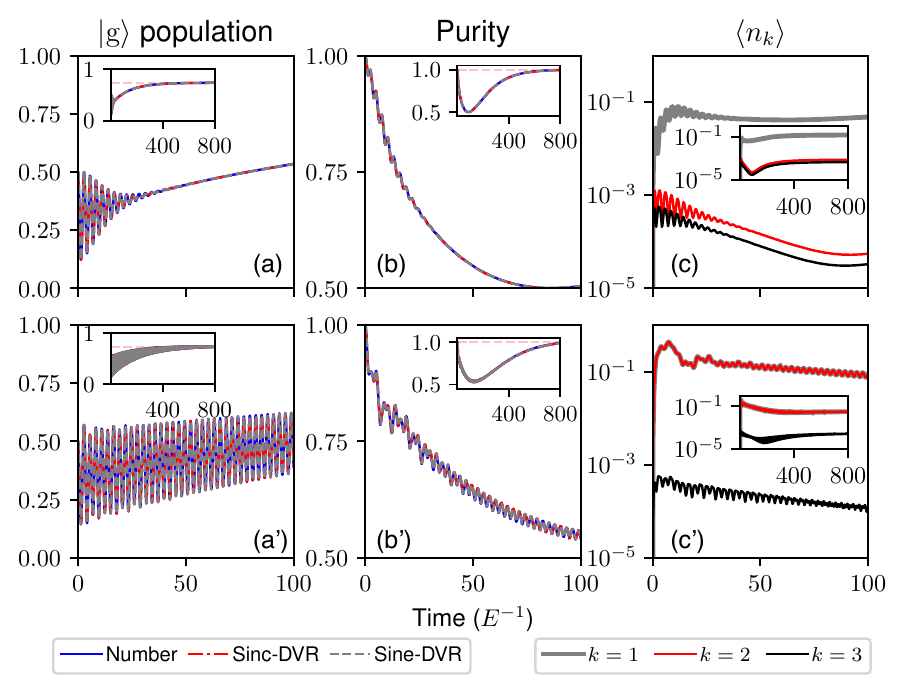}
    \caption{Identical calculation as \fig{fig:dynamics-pd} except that the Hamiltonian of the biased system is set as $\Delta = V= E$ in the number and the position representation (Sinc-DVR and Sine-DVR). The pink dashed line in the insets (a-b, a'-b') corresponds to the ground state of $H_{\text{S}}$.}
\label{fig:dynamics}
\end{figure} 

\Fig{fig:dynamics} shows the dynamics of the qubit model beyond the pure-dephasing limit with $V = \Delta = E$.
This is referred as the biased system case as $\Delta \neq 0$~\cite{Sayer2023}.
In this case, the dephasing is accompanied by relaxation processes.
The parameters of the model correspond to a complex case where the dynamics of the system and environment do not have a clear separation of time scales.
The top panels (a-c) are for the DL bath while the bottom panels (a'-c') for the Brownian bath. 
In both cases, the population of $\ket{\text{g}}$ exhibits Rabi oscillations that decay due to decoherence. This process leads to a reduction of purity to a maximally mixed state ($\mathcal{P}=1/2$). At longer time scales, there is a recovery of purity as the system thermally relaxes to the ground state (insets). The dynamics correctly captures both the early time dynamics~\cite{Gu2017a} and the asymptotic thermal state, and is representative of what is expected of open quantum dynamics. 
Note that different representations (position and number) lead to identical system dynamics, as expected from \eq{eq:eom}. 

With respect to the bexcitonics, for both DL and Brownian bath, the relative bexciton populations indicate that the high-temperature bexciton dominates the dynamics. However, the low-temperature bexcitons are still required to achieve correct thermalization.  
The fact that the bexciton population is non-zero reflects that the system-bath state is not separable at thermal equilibrium. 
 
Changing representations can vastly change the convergence properties of \eq{eq:eom} but leaves bexcitonic properties invariant. We find that the number representation is often more efficient as it employs the exact eigenstates $\ket{\vec{n}}$. Changing the metric changes both the convergence and bexcitonic properties of \eq{eq:eom}, and can be used to develop optimal EQMEs. The relative bexcitonic populations (but not their absolute values) are indicative of the importance of a given bexciton during the dynamics and can be used to test the completeness of \eq{eq:cexpansion} and the numerical convergence of \eq{eq:eom}.

\subsection{Relaxation dynamics of unbiased ($\Delta = 0$) qubit}
\begin{figure}[tb]
    \centering
    \includegraphics[width=\linewidth]{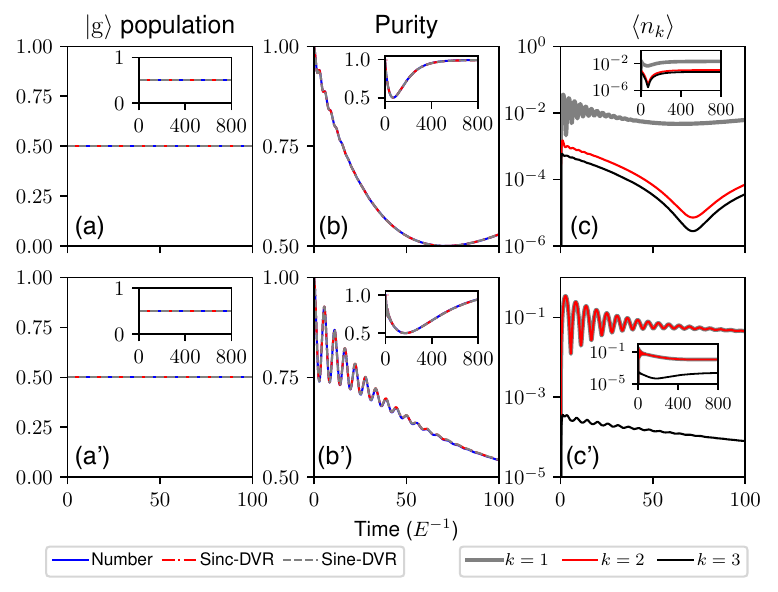}
    \caption{Identical calculation as \fig{fig:dynamics-pd} except that the Hamiltonian of the unbiased system is set as $V = E$ and $\Delta = 0$.}
\label{fig:dynamics-re}
\end{figure} 
\Fig{fig:dynamics-re} shows the dynamics for the qubit with $\Delta=0$ and $V=E$.
This is referred as the unbiased system case~\cite{Sayer2023}.
The top panels (a-c) are for the DL bath while the bottom panels (a'-c') for the Brownian bath. 
In this case, the population for $\ket{\text{g}}$ is fixed at $1/2$, as the rate of population exchange $\ket{\text{g}}\to\ket{\text{e}}$ equals to the rate $\ket{\text{e}}\to\ket{\text{g}}$.
However, the overall purity dynamics shows similar behavior as in the biased case in \fig{fig:dynamics}. That is, the decoherence process leads to a reduction of purity to a maximally mixed state first, and at longer time scales, there is a recovery of purity as the system thermally relaxes and re-purifies to a global steady state.
For the unbiased system we observe that the dynamics of bexcitons shows similar trend as in the biased system for the longer time scale dynamics. 

{\subsection{Influence of the metric on numerical convergence}}

\begin{figure}[tb]
    \centering 
    \includegraphics[width=\linewidth]{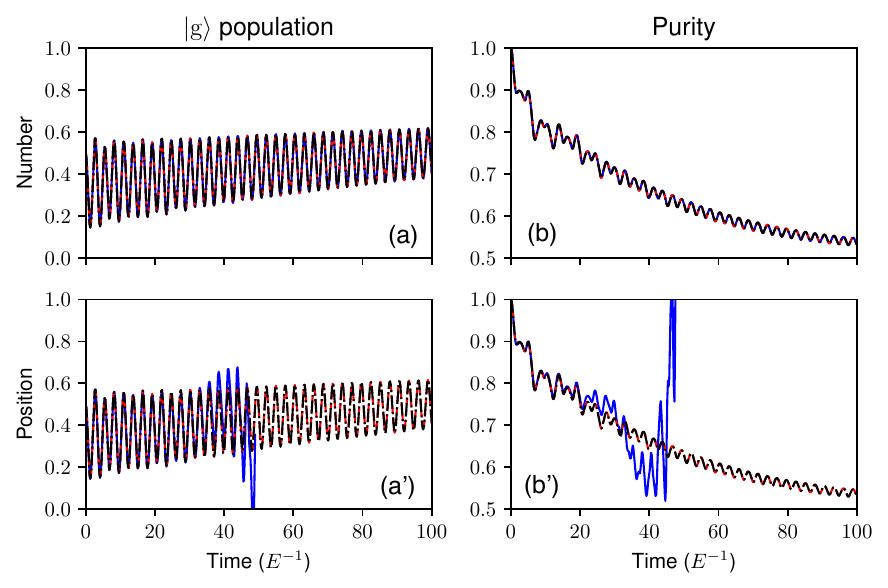}
    \caption{Influence of the metric on the convergence of the number (top panels) and position (bottom panels) representation. The plots show population and purity dynamics for a qubit coupled to a Brownian oscillator with no low-temperature corrections using different metrics for (a-b) number basis and (a'-b') SincDVR basis. We employ as metrics $z_1 = z_2 = \iu$ (blue line), $z_{1}= z_0$  and $z_{2}=-z_0$ where $z_0 = \iu\sqrt{\Re(c_1 + \bar{c}_1)/2} = \iu\sqrt{\Re(c_2 + \bar{c}_2)/2} \approx 0.3 \iu$ (red line), and $z_1=z_2=\iu\sqrt{(\abs{c_1} + \abs{\bar{c}_1})/2} = \iu\sqrt{(\abs{c_2} + \abs{\bar{c}_2})/2} \approx 0.3\iu$ (black line). Other parameter settings are the same as in Fig.~3(a'-b'). Note that while the number basis is largely insensitive to the metric, the convergence in position representation is highly sensitive to $z_k$.} 
    \label{fig:dvr-metric}
\end{figure}

{Changing the metric leaves the open quantum dynamics invariant, but can importantly change its numerical convergence properties. For example, \fig{fig:dvr-metric} shows the dynamics when the qubit is coupled to a Brownian oscillator for varying metrics illustrating the invariance of the results to the choice of metric. However, while the numerical convergence in the number basis is found to be largely insensitive to the metric (top panels, colored lines), the convergence in position representation is highly sensitive to $z_k$ (bottom panels). As shown, in position representation the simple choice of $z_k = \iu$ makes the dynamics to become quickly unstable.  By contrast, the choice  $z_k = \pm\iu\sqrt{\Re(c_k + \bar{c}_k)/2}$ or $z_k=\iu\sqrt{(\abs{c_k} + \abs{\bar{c}_k})/2}$ keep the dynamics stable for the times shown. 
We find that larger $\abs{{z}_k}$ does not necessarily suppresses the divergence of auxiliary density matrices and $\ev{n_k}$, as in \eq{eq:dissipator} both $\hat{z}_k$ and $\hat{z}_k^{-1}$ occurs in the expression. That is, by suppressing the term associated with the creation operator $\hat{\alpha}^\dagger$ we also enlarge the term associated with the annihilation operator, which amplifies the error from truncating at finite depth. Overall, these results suggest that the metric can be considered as a simulation parameter that can be optimized to achieve enhanced numerical convergence.}


{\section{Applications}\label{stn:applications}}

{We now show how the bexcitonic approach can lead to useful insights, and to the development of  efficient strategies to propagate the quantum dynamics. }

{\subsection{Bexcitonic perspective of the numerical instability in the HEOM}}
\begin{figure}[tb]
    \centering 
    \includegraphics[width=\linewidth]{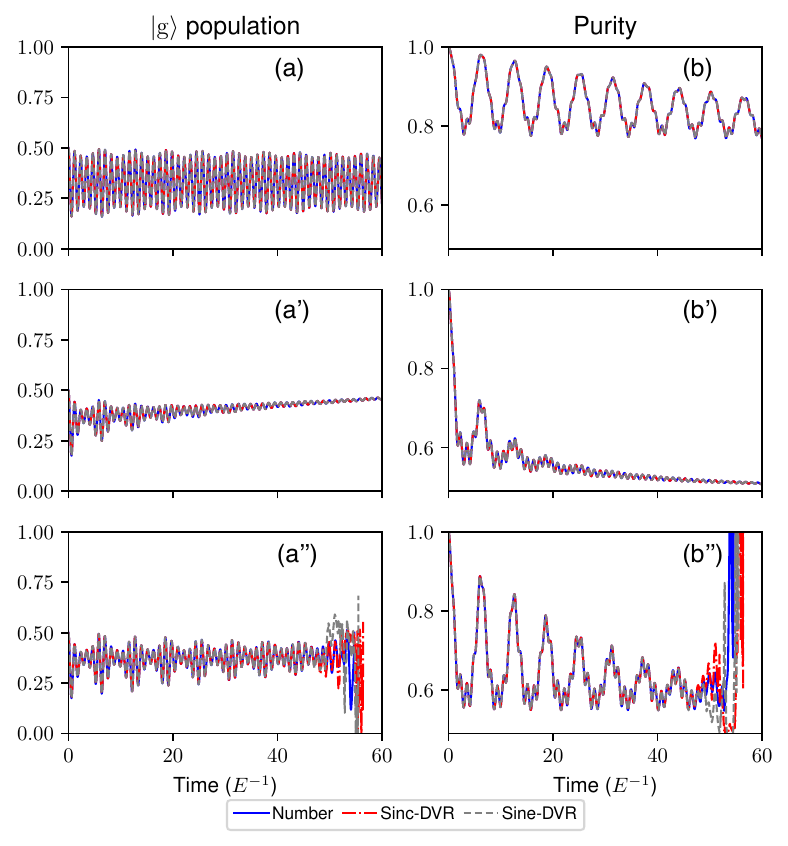} 
    \caption{Numerical instability of \eqs{eq:eom}{eq:dissipator} for a qubit ($\Delta=5E$ and $V = E$)  coupled to a Brownian oscillator with damping rate $\eta$ and coupling $\lambda$ in number and position representation. The plots show the population and purity dynamics for (a, b) weak coupling ($\lambda = 0.2 E$) and long correlation time ($\eta = 0.01 E$);  (a', b') strong coupling ($\lambda = E$) and short correlation time ($\eta = 0.05 E$); and  (a'', b'') strong coupling ($\lambda = E$) and long correlation time ($\eta = 0.01 E$).  For last case, weak damping rate  together with strong coupling results in instability for the dynamics in all representations. All simulations use a depth of $40$ and metric $z_k = \iu\sqrt{\Re(c_k + \bar{c}_k)/2}$.}
    \label{fig:brownian-strong}
\end{figure}

\begin{figure}[tb]
    \centering 
    \includegraphics[width=\linewidth]{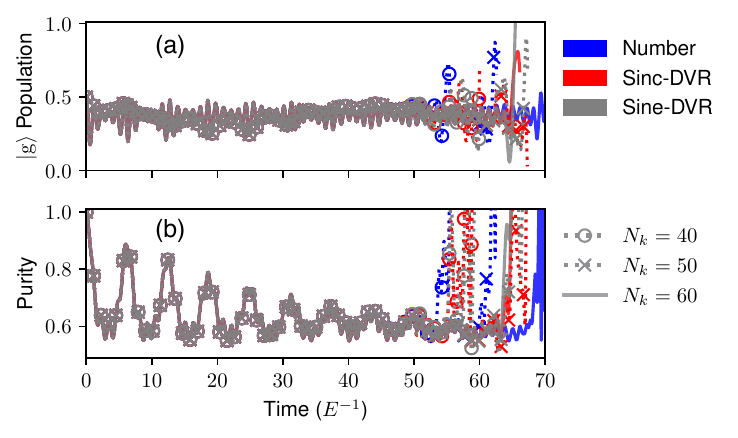}
    \caption{Effect of changing the HEOM depth $N_k$ on the convergence in both number and position representation.  Dynamics of (a) population of $\ket{\mathrm{g}}$ and (b) purity for the qubit under simulation conditions identical to those in Fig.~\ref{fig:brownian-strong}(a'') and (b'') but with varying $N_k$.  Increasing $N_k$ from 40 to 60 only moderately extends the convergence range of HEOM.}%
    \label{fig:tiers}
\end{figure}

\begin{figure}[tb]
    \centering
    \includegraphics[width=\linewidth]{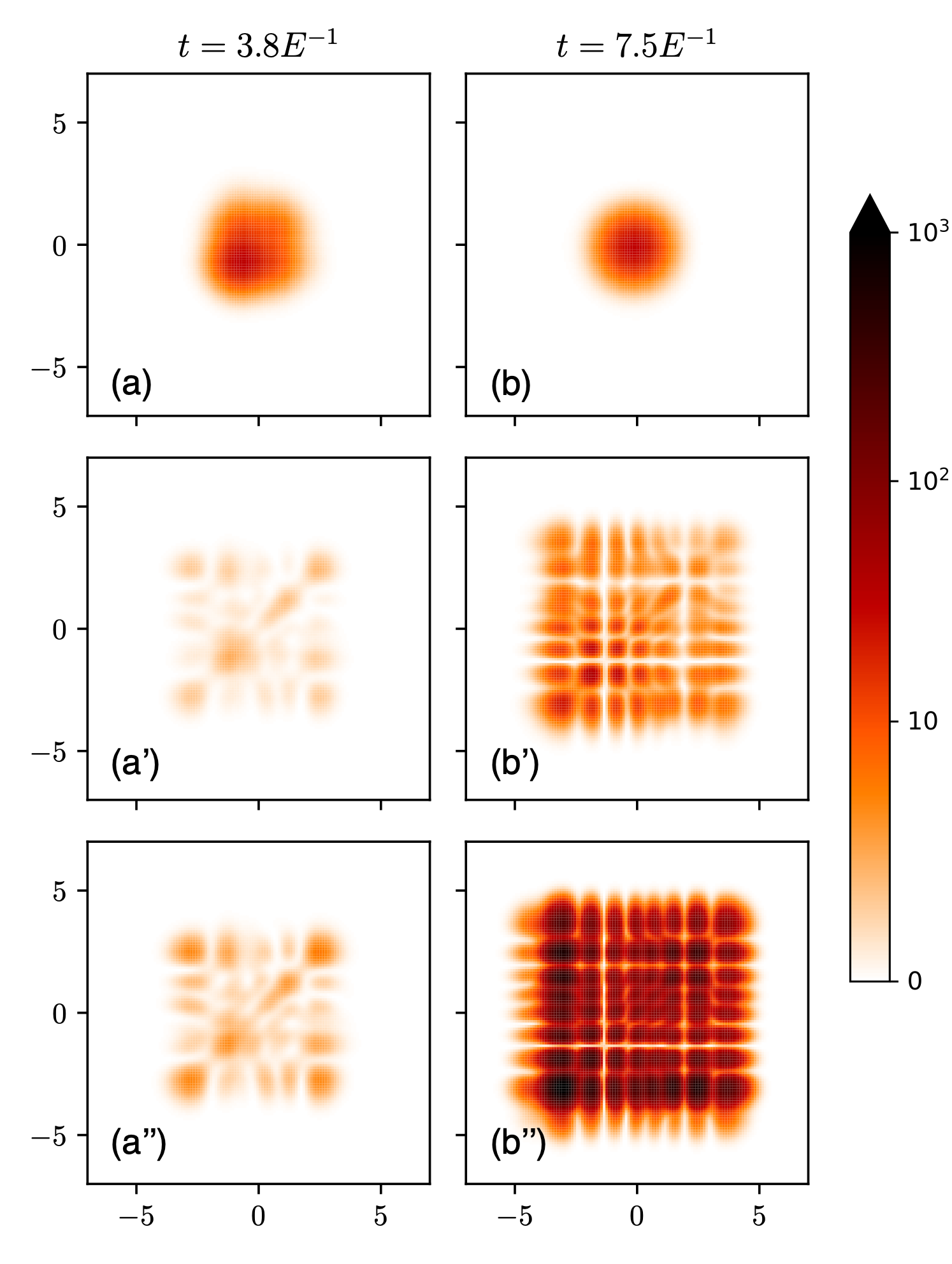} 
    \caption{Numerical instability of the HEOM as seen from bexcitonic density distribution. The plots show the $||\varrho(x_1, x_2, t) ||^2$ distribution at $t = 3.8E^{-1}$ (left column) and $t = 7.5E^{-1}$ (right column). (a-b) weak coupling ($\lambda = 0.2 E$) and long correlation time ($\eta = 0.01 E$);  (a'-b') strong coupling ($\lambda = E$) and short correlation time ($\eta = 0.05 E$); and  (a''-b'') strong coupling ($\lambda = E$) and long correlation time ($\eta = 0.01 E$). The instability of the dynamics is due to the population of highly excited bexcitons that lead to a large spatial structure for $||\varrho(x_1, x_2, t)||^2$. In (a-b) $||\varrho(x_1, x_2, t)||^2$ is scaled by $10^{2.4}$ for clarity.}
    \label{fig:vis}
\end{figure} 

{A numerical instablity in HEOM has been reported recently, especially in the case of the weakly-damped Brownian oscillator~\cite{Dunn2019, Yan2020}.
To test this further, we have performed  simulation of the qubit ($\Delta=5E$ and $V = E$) when the reorganization energy is relatively small ($\lambda=0.2E$) or large  ($\lambda=E$), and when the bath damping rate ($\eta$) is relatively large ($\eta=0.05E$) or  small ($\eta=0.01E$) for the Brownian oscillator bath. The results, shown in \fig{fig:brownian-strong}, show that the instability occurs in both the number and position basis. For the specific metric chosen, the instability develops at about the same point of the dynamics.}

{As shown in \fig{fig:tiers}, by using a larger depth $N_k$ the convergence of the dynamics can be achieved in a longer range of time.
For example, for the simulations in Fig.~\ref{fig:brownian-strong}(a'') and (b''), increasing the depth $N_k$ from 40 to 60 improves the range of time of convergence from $\sim 50E^{-1}$ to $\sim 63E^{-1}$.
However, the actual computational time per step increases by a factor of $\sim 5$ in both position and number basis in a 8-core Intel Xeon Gold 6330 CPU node. For this reason, in practice, HEOM computations require bexciton space truncation that introduces such instability in the equations of motion for the strong coupling and small bath damping rate case.}

The origin of this instability has been previously analyzed by mathematically investigating the distribution of eigenvalues for the propagator~\cite{Dunn2019, Yan2020}.  
Here we provide an intuitive explanation based on the bexcitons.

{\Fig{fig:vis} shows the bexcitonic density distribution  $||\varrho{(x_1,x_2,t)}||^2 = \ip{\varrho(t)}{x_1 x_2} \ip{x_1 x_2}{\varrho(t)}$ in position representation at $t = 3.8E^{-1}$ and $ 7.5E^{-1}$ which is early in the dynamics and before the onset of the instability. In the computations, we do not include the low-temperature corrections and therefore two bexcitons with coordinates $x_1$ and $x_2$ are needed. \fig{fig:vis}(a-b) correspond to the dynamics in \fig{fig:brownian-strong}(a-b) and that in \fig{fig:vis}(a'-b') to the one in \fig{fig:brownian-strong}(a'-b'). Both these cases are numerically stable in the range of time investigated. In turn, \fig{fig:vis}(a''-b'') correspond to the numerically unstable dynamics in \fig{fig:brownian-strong}(a''-b''). In all cases, at initial time $||\varrho{(x_1,x_2,t)}||^2$ is a localized Gaussian function centered at $(0, 0)$. The degree of deviation of $||\varrho{(x_1,x_2,t)}||^2$ from the initial Gaussian is a consequence of the system-bath interaction. As seen, the reason why the dynamics becomes unstable is because as the coupling strength and the bath correlation time increases, the bexcitonic state acquires strong spatial structure which is numerically challenging to capture.  In fact, the unstable case leads to a highly structured bexcitonic state with extensive ripples and nodes. 
These dramatic oscillations and large magnitude of the bexcitonic density distribution in position space lead to large partial derivatives in \eq{eq:pde-BME}. 
This phenomena make the HEOM and other bexcitonic methods better at simulating systems with weaker coupling to the bath and stronger damping in the bath dynamics.
This also results in more auxiliary density matrices needed to capture these highly excited bexcitonic states with many nodes. Thus, the instability of the HEOM can be seen as emerging because of the population of highly excited bexcitons. By contrast, numerically stable dynamics is observed when there is enough dissipation in the bath dynamics to mitigate the population of highly excited bexcitonic states.}

{Ref.~\onlinecite{Ikeda2022}  claims that the coordinate representation of the HEOM is more stable and efficient than the original HEOM theory and demonstrated this in the context of a vibronic system (two electronic levels plus one vibration) coupled to a Drude-Lorentz environment. By contrast, we find that this is not necessarily a general observation as the numerical convergence of the number representations is seen to be superior than the coordinate representation in our numerical examples. Further, for Brownian oscillator environments both number and position representation show divergence in the propagation due to the population of highly excited bexcitons when the  damping rate of the bath correlation function is small and the system-bath coupling is strong.}

{\subsection{Mode-combination of the bexcitons}}

\begin{figure}[tb]
    \centering 
    \includegraphics[width=\linewidth]{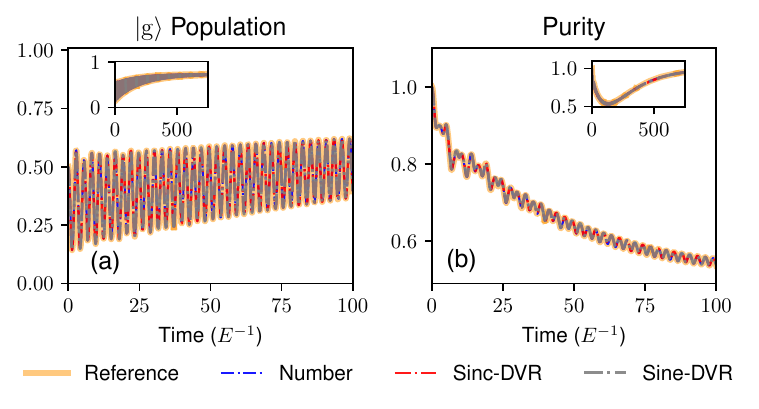}
    \caption{Mode-combination generated by compressing all bexcitons to a single effective one that tracks the influence of the bath. The plots show the population and purity dynamics for the system with $\Delta=E$ and $V = E$ coupled to a Brownian oscillator with $K=4$ features with and without mode-combination. 
    The reference corresponds to computations using the usual number basis without mode-combination. The depth $N_k=10$ for the number basis and 40 for position.
    The number of SPFs is $10$ for all mode-combined cases.}
    \label{fig:mc}
\end{figure}

{In the bexciton approach, the influence of the environment is exactly captured through $\ket{\varrho(t)}$ that include both the physical system $\rho_\text{S}(t)$ and the bexcitons. The main limitation of this approach is that the dimensionality of the space for representing $\ket{\varrho(t)}$ grows exponentially with the number of features of the bath $K$, and hence,
the computational memory requirements of the method quickly become intractable as the number of features $K$ grows. 
Notice, however, that the non-uniqueness of \eq{eq:bcf} suggests the dimensionality of the bexciton space is greater  than what is actually needed for tracking the open quantum dynamics suggesting that it is possible to compress the dynamics.}

As an example, with the help of the bexciton quasiparticle picture one can incorporate the mode-combination technique developed for the MCTDH method~\cite{Meyer1990,Wang2003} into the HEOM {in both number and coordinate representation}. 
{Mode-combination has been proposed \cite{Yan2021, Ke2023} to enhance the computational efficiency of the usual HEOM. 
This technique corresponds to the tensor tree extension of the HEOM. 
Here we show that the validity of this technique in HEOM can be interpreted as a mode-combination of the bexcitons and use the bexcitonic structure to introduce mode-combination in coordinate representation. This is an example of how \eqs{eq:eom}{eq:dissipator} enables translating technical advances from one HEOM variant to others.}

Specifically,  we introduce a set of orthonormal single-particle bexciton functions (SPFs) {$\ket{\chi_\sigma(t)} = \sum_{n_1 \ldots n_K}{C^{(\sigma)}_{n_1\ldots n_K}(t)}\ket{\vec{n}}$} that combines all bexcitons with 
$\ip{\chi_\sigma(t)}{\chi_{\sigma'}(t)} = \delta_{\sigma \sigma'}.$
Using them, the EDO can be expressed as  
$
    \ket{\varrho(t)} = \sum_{\sigma}  \ket{\varrho^{\text{C}}_{\sigma}(t)}\ket{\chi_\sigma(t)},
$
where $\ket{\varrho^{\text{C}}_{\sigma}(t)} = \sum_{ij}\ket{i} R^{(\sigma)}_{ij}(t)\bra{j}$ contains the auxiliary density matrices that correspond to the combined bexciton $\ket{\chi_\sigma(t)}$ with collective index $\sigma$, and $R_{ij}^{(\sigma)}(t)$ are the matrix elements. 
The idea is to isolate a few SPFs that capture the main dynamics due to all $K$ bexcitons. 
For this, we follow the multi-layer MCTDH strategy
\cite{Meyer2018,Wang2018on} to propagate each $\ket{\chi_\sigma(t)}$ for a given chosen level of compression by controlling the number of SPFs taken into account.



To show the advantage of such {bexciton} mode-combination strategy, we performed the calculation for the system with $\Delta=E$ and $V = E$ coupled to a Brownian oscillator with 2 low-temperature correction terms ($K=4$) with or without the mode-combination strategy. Other parameter settings are the same as in \fig{fig:dynamics}(a'-b'). The results are shown in Fig.~\ref{fig:mc}.
{By employing the mode combination with the number of SPFs to be $10$, the memory requirements to store the combined state of the system and environment at a given time is reduced from $640$~KB to $49$~KB for $N_k=10$. }

{Using \eq{eq:eom} we extend these technical advances to coordinate representation where  
${\varrho(\vec{x}, t)} = 
    \sum_{\sigma}  (\varrho^{\text{C}}_{\sigma}(t))\chi_\sigma(\vec{x},t),
$
where ${\chi_\sigma(\vec{x},t)} $ is the mode-combination of bexciton function in position  representation $\vec{x}$ with given $\sigma$ at time $t$.
With the DVR basis, we can further represent
$\chi_\sigma(\vec{x},t)= \sum_{m_1 \ldots m_K}{C^{(\sigma)}_{m_1\ldots m_K}(t)} {\phi_{m_1}(x_1)} \cdots {\phi_{m_K}(x_K)}$ where ${\phi_{m_k}(x_k)}$ is the DVR basis function at $m_k$-th grid point for the $k$-bexciton. 
Each $\chi_\sigma(\vec{x},t)$ can further be propagated using the MCTDH strategy in the DVR basis. 
\Fig{fig:mc} shows results obtained from mode combination in position representation, yielding identical dynamics to that obtained in number representation.  
In this case, the mode combination reduces the memory requirements to store an EDO from $164$~MB to $0.53$~MB for the depth chosen as $N_k=40$.
}

{
For the number representation with $N_k=10$, which is an efficient way to propagate the dynamics of our models, we find that the mode-combination offers a $\sim 1.5\times$ speedup in a 8-core Intel Xeon Gold 6330 CPU node. 
In turn, for position representation and $N_k=40$ the speedup is $\sim 200\times$  using the Sinc-DVR basis and $\sim 300\times$ for the Sine-DVR basis, suggesting a high level of redundancy. }


\section{Conclusion}
\label{stn:conclusions}

In conclusion, we have {developed a quasiparticle approach for the exact open quantum dynamics of systems in interaction with bosonic thermal environments of arbitrary complexity based on a generalization of the Hierarchical Equations of Motion (HEOM)}. 
In this approach, the dynamics is exactly captured by the quantum system interacting with a few bexcitons, fictitious bosonic quasiparticles each one arising from a distinct feature of the bath correlation function. {Bexciton creation and annihilation connect the auxiliary density matrices in the HEOM. Since bexcitonic operators can be represented in different bases and are associated with a tunable metric, the approach enables the straightforward and systematic development of HEOM variants. Because all these HEOM variants are seen to be specific realizations of the same bexcitonic equations of motion, when converged, they yield the same dynamics for a given decomposition of the bath correlation function. Further, if technical advances in one variant is made, it can immediately inspire related advances in other variants.  However, we find that the convergence property and numerical stability may vary because of the different errors introduced in the truncation of bexcitonic basis in different realizations. Thus, the bexcitonic basis and metric can be chosen to optimize the convergence properties of the dynamics.}

We implemented these equations both in number and position representation, showed that they were numerically stable and made the code publicly accessible. While bexcitonic properties are unphysical, they can be used to monitor numerical convergence and guide the development of convenient and computationally efficient exact quantum master equations.
{As an example, we used this feature to explain the origin of the instability of the HEOM when the bath is a weakly damped Brownian oscillator and show that it leads to the strong population of highly excited bexcitons. Further, by taking advantage of the particle-like features of the bexcitons, we introduced the concept of mode-combination of physical degrees of freedom developed in multi-configuration time-dependent Hartree and applied to the bexcitons for a more efficient propagation of the dynamics {in both number and position representation}.}

{Future prospects include extending the theory to fermionic/spin environments and non-linear coupling, determining strategies to optimize the metric and representation, and introducing general tensor network decomposition and corresponding algorithms to enhance the computational efficiency of the method.}

\begin{acknowledgments}
    This work was supported by a PumpPrimer II award of the University of Rochester and, partially, by the National Science Foundation under Grant Nos.~CHE-2102386 and PHY-2310657.
    Computing resources are provided by the Center for Integrated Research Computing at the University of Rochester. The authors thank Gabriel Landi and Oliver K\"uhn for very helpful discussions on the subject.
\end{acknowledgments}

\appendix

\begin{widetext}
    
\section{Derivation of the bexcitonic exact quantum master equation \eqs{eq:eom}{eq:dissipator}}\label{sec:derivation}

To capture the open quantum dynamics exactly, we need to take into account how each $\tilde{f}_k$ in \eq{eq:propagator} influences the system's dynamics. 
The master equation is derived by taking the time-derivative of \eq{eq:adm-def} as
\begin{multline} \label{eq:si-diff} 
    \pdv{t} \tilde{\varrho}_{\vec{n}}(t) 
    =
    \sum_{k=1}^K
    \mathcal{T} 
    \frac{1}{Z_k(n_k)\sqrt{n_k!}}
    \qtya{\pdv{t} \tilde{f}_k^{n_k}(t, 0)}
    \qtya{
        \prod_{j\neq k}\frac{\tilde{f}^{n_j}_j(t, 0)}{ Z_{j}(n_j)\sqrt{n_j!}}
    }
    \tilde{\mathcal{F}}(t, 0) \rho_\text{S}{(0)} 
    \\
    + \mathcal{T} 
    \qtya{\prod_{k=1}^K \frac{\tilde{f}^{n_k}_k(t, 0)}{Z_k(n_k)\sqrt{n_k!}}
    }
    \qtya{\pdv{t}\tilde{\mathcal{F}}(t, 0)} \rho_\text{S}{(0)}
    \equiv \sum_{k=1}^K A_k + A_0
    . 
\end{multline}
Note that the derivatives of the bexciton generator $\tilde{f}_k$ are
\begin{equation}
    \begin{aligned}
        \pdv{t} \tilde{f}_k(t, 0)
        &= c_k\tilde{Q}_\text{S}^>(t) -\bar{c}_k\tilde{Q}_\text{S}^<(t)
        + \sum_{k'}  \gamma_{kk'} \tilde{f}_{k'}(t, 0).
    \end{aligned}
\end{equation}
Here we have used \eq{eq:bcf-basis} and
\begin{equation} 
        \pdv{t} \tilde{\theta}_k(t, 0) 
        =
        \tilde{Q}_\text{S}(t)\psi_k(t-t)
        +
        \int_0^{t} \tilde{Q}_\text{S}(u)\pdv{t} \psi_k(t-u) \dd{u}
        = \tilde{Q}_\text{S}(t)
        + \sum_{k'} \gamma_{kk'} \tilde{\theta}_{k'}(t, 0). 
\end{equation}
Using these results, each term in the first sum in \eq{eq:si-diff} becomes
\begin{multline}
\label{eq:diff1}
    A_k
    =
    {z_{k, n_k}^{-1}} \sqrt{n_k}\qtya{c_k\tilde{Q}_\text{S}^>(t) -\bar{c}_k\tilde{Q}_\text{S}^<(t)} \tilde{\rho}_{\vec{n} - \vec{1}_k}(t)
    + n_k \gamma_{kk} \tilde{\rho}_{\vec{n}}(t) 
    + \sum_{k'\neq k} {z_{k, n_k}^{-1}}  z_{k', n_{k'}+1}  \sqrt{n_k(n_{k'}+1)} \gamma_{kk'} 
        \tilde{\varrho}_{\vec{n} - \vec{1}_{k} + \vec{1}_{k'} }(t)
\end{multline} 
for $k = 1,\ \ldots,\ K$. Using  \eq{eq:if}, the last part in \eq{eq:si-diff} can be expressed as
\begin{equation}
    A_0
    =
    - \mathcal{T}
    \qtya{ 
    \prod_{j=1}^K  \frac{\tilde{f}^{n_j}_j(t, 0)}{Z_j(n_j)\sqrt{n_j!}}
     } \sum_{k=1}^K \tilde{Q}^{\times}_\text{S}(t) \tilde{f}_k(t, 0)
    \tilde{\mathcal{F}}(t, 0)\rho_\text{S}{(0)}.
\end{equation}
Hence,
\begin{equation}\label{eq:diff2}
    A_0
    = - \sum_{k=1}^K z_{k, n_{k}+1} \sqrt{n_k + 1}\tilde{Q}^{\times}_\text{S}(t)
    \tilde{\varrho}_{\vec{n} + \vec{1}_k}(t)
    ,
\end{equation} 
where we have used the fact that
\begin{equation}
    \tilde{\rho}_{\vec{n} \pm \vec{1}_k}(t) =
    \mathcal{T}  \frac{\tilde{f}^{n_k \pm 1}_k(t, 0)}{ Z_{k}(n_k \pm 1) \sqrt{(n_k \pm 1)!}}
        \qtya{\prod_{j\neq k} 
        \frac{\tilde{f}^{n_j}_j(t, 0)}{Z_j(n_j)\sqrt{n_j!}}}
        \tilde{\mathcal{F}}(t, 0) \rho_\text{S}{(0)},
\end{equation}
and
\begin{equation} 
    \tilde{\varrho}_{\vec{n} - \vec{1}_{k} + \vec{1}_{k'} }(t) =
    \mathcal{T}
    \frac{\tilde{f}^{n_k - 1}_k(t, 0)}{Z_k({n_k - 1})\sqrt{({n_k - 1})!}}
     \frac{\tilde{f}^{n_{k'} + 1}_{k'}(t, 0)}{Z_{k'}(n_{k'} + 1)\sqrt{(n_{k'} + 1)!}}
    \qtya{\prod_{j\not\in \{k,k'\}}   
    \frac{\tilde{f}^{n_j}_j(t, 0)}{Z_j({n_j})\sqrt{{n_j}!}}
    } 
        \tilde{\mathcal{F}}(t, 0) \rho_{\text{S}}^{(0)}
\end{equation}
for $k'\neq k$.
From Eqs.~\eqref{eq:si-diff}, \eqref{eq:diff1} and \eqref{eq:diff2}, 
\begin{multline}\label{eq:heom-i}
        \pdv{t} \tilde{\varrho}_{\vec{n}}(t) 
        = \sum_k 
        n_k \gamma_{kk} \tilde{\varrho}_{\vec{n}}(t) 
        + 
        \sum_{k'\neq k}  {z_{k, n_k}^{-1}} z_{k', n_{k'}+1} 
        \sqrt{n_k(n_{k'}+1)} \gamma_{kk'} \tilde{\varrho}_{\vec{n} - \vec{1}_{k} + \vec{1}_{k'} }(t) 
        \\
        + \sum_{k} {z_{k, n_k}^{-1}}  \sqrt{n_k} \qtya{
            c_k \tilde{Q}_\text{S}^>(t) - \bar{c}_k \tilde{Q}_\text{S}^<(t) }\tilde{\varrho}_{\vec{n} - \vec{1}_k}(t) 
        - \sum_{k}  z_{k, n_k+1}   \sqrt{n_k + 1}\tilde{Q}^{\times}_\text{S}(t)\tilde{\varrho}_{\vec{n} + \vec{1}_k}(t)
        .
\end{multline}
Thus, to capture the exact open quantum dynamics it is necessary to follow the dynamics of all auxiliary density matrices that define the EDO.  

We define an extended density operator (EDO) as a collection of these auxiliary density matrices. 
We arrange these matrices as a vector of matrices 
$\ket{\tilde{\varrho}(t)} =\sum_{\vec{n}} \tilde{\varrho}_{\vec{n}}(t) \ket{\vec{n}} $ in a basis $\{\ket{\vec{n}}\equiv \ket{n_1}\otimes\cdots\otimes \ket{n_k}\otimes\cdots\otimes\ket{n_K} \}$
such that $\tilde{\varrho}_{\vec{n}}(t) = \ip{ \vec{n} }{ \tilde{\varrho}(t) }$. 
The physical system's density matrix corresponds to $\tilde{\varrho}_{\vec{0}}(t) \equiv  \tilde{\rho}_{\text{S}}(t)$. In this space, we can define the creation $\alpha^\dagger_k$ and annihilation $\alpha_k$ operators associated to the $k$-th feature of the bath such that
\begin{equation}
    \hat{\alpha}^\dagger_k \ket{{n_k}} = \sqrt{n_k + 1} \ket{{n_k+1}},\    \hat{\alpha}_k \ket{{n_k}} = \sqrt{n_k} \ket{{n_k-1}}, 
\end{equation}
and $[\hat\alpha_k, \hat\alpha_{k'}^\dagger] = \delta_{k,k'}$, $\hat{n}_k=\hat\alpha_{k}^\dagger\hat\alpha_{k}$.
We define a metric operator for the $k$-bexciton as
$\hat{z}_k \ket{n_k} =  z_{k,n_k} \ket{n_k}$. 
Hence,
\begin{align} 
\label{eq:up}
    \mel{{\vec{n}
}}{\hat{z}_k^{-1}\hat{\alpha}^\dagger_k }{\tilde{\varrho}(t)} &= {z_{k, n_k}^{-1}} \sqrt{n_k}
\ip{
    {n_1} \cdots (n_k-1) \cdots {n_K}
}{\tilde{\varrho}(t)}
    = {z_{k, n_k}^{-1}}\sqrt{n_k} \tilde{\varrho}_{\vec{n} - \vec{1}_k}(t), \\
\label{eq:down}
    \mel{{\vec{n}
}}{\hat{\alpha}_k \hat{z}_k}{\tilde{\varrho}(t)} &= z_{k, n_k+1}  \sqrt{n_k + 1}
\ip{
    {n_1} \cdots (n_k+1) \cdots {n_K}
}{\tilde{\varrho}(t)}
    = z_{k, n_k+1} \sqrt{n_k + 1} \tilde{\varrho}_{\vec{n} + \vec{1}_k}(t), 
\end{align}
and
\begin{multline}
\label{eq:up-down}
\mel{{\vec{n}}}{\hat{z}_k^{-1}\hat{\alpha}_k^\dagger \hat{\alpha}_{k'}\hat{z}_{k'}}{\tilde{\varrho}(t)} = {z_{k, n_k}^{-1}} z_{k', n_{k'}+1} \sqrt{n_k(n_{k'} + 1)}\ip{{n_1} \cdots (n_k-1)  \cdots (n_{k'}+1) \cdots {n_K}} {\tilde{\varrho}(t)} \\
= {z_{k, n_k}^{-1}} z_{k', n_{k'}+1} \sqrt{n_k(n_{k'} + 1)} \tilde{\varrho}_{\vec{n} - \vec{1}_k + \vec{1}_{k'}}(t).
\end{multline} 
Inserting Eqs.~\eqref{eq:up}, \eqref{eq:down} and \eqref{eq:up-down} into \eq{eq:heom-i}, we obtain
\begin{equation}  
        \pdv{t} \ip{ \vec{n} }{ \tilde{\varrho}(t) }
        = \sum_{kk'} \gamma_{kk'} \bra{\vec{n}} \hat{z}_k^{-1}\hat{\alpha}_k^\dagger \hat{\alpha}_{k'}\hat{z}_{k'} \ket{\tilde{\varrho}(t) }
        - \sum_{k} \tilde{Q}^{\times}_\text{S}(t)
        \bra{\vec{n}} \hat{\alpha}_{k}\hat{z}_{k} \ket{\tilde{\varrho}(t)}
        + \sum_{k} \qtya{
            c_k \tilde{Q}_\text{S}^>(t) - \bar{c}_k \tilde{Q}_\text{S}^<(t) }
           \bra{\vec{n}} \hat{z}_k^{-1}\hat{\alpha}_k^\dagger \ket{\tilde{\varrho}(t)}
        .  
\end{equation}
Since the the equation must be valid for arbitrary $\ket{\vec{n}}$, then
\begin{equation}
\label{eq:eomtmp}
        \pdv{t}  \ket{\tilde{\varrho}(t) }
        = \sum_{kk'} \gamma_{kk'} \hat{z}_k^{-1}\hat{\alpha}_k^\dagger \hat{\alpha}_{k'}\hat{z}_{k'} \ket{\tilde{\varrho}(t) }
        - \sum_{k} \tilde{Q}^{\times}_\text{S}(t)
         \hat{\alpha}_{k}\hat{z}_{k} \ket{\tilde{\varrho}(t)} 
         + \sum_{k} \qtya{
            c_k \tilde{Q}_\text{S}^>(t) - \bar{c}_k \tilde{Q}_\text{S}^<(t) }
             \hat{z}_k^{-1}\hat{\alpha}_k^\dagger \ket{\tilde{\varrho}(t)}
        . 
\end{equation} 

To obtain the final EQME we only need to express \eq{eq:eomtmp} in the Schr\"odinger picture. Since the interaction picture is only for the physical system, and not for the introduced bexcitons, the procedure just requires changing the system operators to the Schr\"odinger and recovering the systematic dynamics due to the system's Hamiltonian. In the Schr\"odinger picture,
\begin{equation}\label{eq:si-bme}
    \pdv{t} \ket{{\varrho}(t)} =\qtya{ - {\iu} {H}_\text{S}^{\times} + \sum_{k=1}^{K}\mathcal{D}_k} \ket{{\varrho}(t)},
\end{equation}
where 
\begin{equation}
    \label{eq:si-dissipator}
       \mathcal{D}_k 
       = \sum_{k'} \gamma_{kk'} \hat{z}_{k}^{-1} {\hat{\alpha}}^\dagger_k {\hat{\alpha}}_{k'} \hat{z}_{k'} +  \qtya{
        c_k {Q}_\text{S}^> - \bar{c}_k {Q}_\text{S}^< }
        \hat{z}_{k}^{-1} \hat{\alpha}^\dagger_k 
       - {Q}^{\times}_\text{S}
        \hat{\alpha}_k\hat{z}_{k}
\end{equation}
are the dissipators in the dynamics.
This yields \eqs{eq:eom}{eq:dissipator} in the main text.

\end{widetext}

\bibliography{main.bib}

\end{document}